\documentclass[%
 aip,
 amsmath,amssymb,
 reprint,%
]{revtex4-1}
\usepackage{graphicx}
\usepackage{dcolumn}
\usepackage{bm}
\usepackage[utf8]{inputenc}
\usepackage[T1]{fontenc}
\usepackage{mathptmx}
\usepackage{etoolbox}
\usepackage{xcolor}
\colorlet{Mycolor}{black}
\makeatletter
\def\@email#1#2{%
 \endgroup
 \patchcmd{\titleblock@produce}
  {\frontmatter@RRAPformat}
  {\frontmatter@RRAPformat{\produce@RRAP{*#1\href{mailto:#2}{#2}}}\frontmatter@RRAPformat}
  {}{}
}%
\makeatother
\begin{document}

\preprint{AIP/123-QED}

\title[Speckle-based 3D sub-diffraction imaging through a multimode fiber]{Speckle-based 3D sub-diffraction imaging through a multimode fiber}
\author{Zhouping Lyu}
\email{z.lyu@arcnl.nl}
\affiliation{ 
Advanced Research Center for Nanolithography (ARCNL), Science Park 106, 1098 XG Amsterdam, The Netherlands
}
\affiliation{
Department of Physics and Astronomy, and LaserLaB, Vrije Universiteit, De Boelelaan 1081, 1081 HV Amsterdam, The Netherlands
}

\author{Shih-Te Hung}%
\affiliation{Delft Center for Systems and Control, Delft University of Technology, Delft, The Netherlands}

 \author{Carlas S. Smith}
\affiliation{Delft Center for Systems and Control, Delft University of Technology, Delft, The Netherlands}
\affiliation{Department of Imaging Physics, Delft University of Technology, Delft, The Netherlands}

\author{Lyubov V. Amitonova}
\affiliation{ 
Advanced Research Center for Nanolithography (ARCNL), Science Park 106, 1098 XG Amsterdam, The Netherlands
}
\affiliation{
Department of Physics and Astronomy, and LaserLaB, Vrije Universiteit, De Boelelaan 1081, 1081 HV Amsterdam, The Netherlands
}

\date{\today}

\begin{abstract}
A flexible multimode fiber is an exceptionally efficient tool for \emph{in vivo} deep tissue imaging. Recent advances in compressive multimode fiber sensing allow for imaging with sub-diffraction spatial resolution and sub-Nyquist speed. At present, the technology is limited to imaging in a two-dimensional (2D) plane near the fiber distal facet while in real applications it is very important to visualize three-dimensional (3D) structures. Here we propose a new approach for fast sub-diffraction 3D imaging through a multimode fiber by using a single 2D scan, speckle illumination, and bucket detection. We experimentally demonstrate precise image plane location as well as 3D imaging of samples with various scattering coefficients. The full width at half maximum (FWHM) of the point spread function along the axial direction is 3 times smaller than the diffraction limit. Our study grants depth-resolving capacity to ultra-thin super-resolution fiber endoscopes for life science and medical applications.
\end{abstract}

\maketitle

\section{Introduction}

Breakthroughs in biology and life sciences are driven by continuous progress in imaging techniques. \textcolor{Mycolor}{Fluorescence imaging has been crucial in this progress, allowing for specific labeling, localization, and visualization of structures of cells and processes with high specificity and sensitivity \cite{ntziachristos2003fluorescence,lauwerends2021real}.} Modern super-resolution methods of optical microscopy enable non-destructive imaging with high subcellular spatial resolution \cite{galbraith2011super,hell20152015}. 
Despite its remarkable capabilities, state-of-the-art super-resolution microscopy has very shallow penetration depth allowing for imaging of only superficial layers, up to $120~\mu$m deep at best~\cite{urban2011sted,jing2021super}. Moreover, the higher resolution system requires a higher numerical aperture (NA), which makes the point spread function (PSF) more susceptible to scattering. This results in a further reduced penetration depth. In contrast, micro-endoscopes excel at deep imaging as a minimally invasive probe due to the inherently small footprint~\cite{andresen2016ultrathin,badt2022real}.

Multimode fibers (MMFs) support the propagation of thousands of spatial modes providing the highest information density for the given footprint, which is ideal for minimally invasive endoscopy. Achieving higher resolution can be as straightforward as using a high-NA fiber, without compromising the penetration depth of the fiber. However, the information is scrambled by mode mixing and modal dispersion\cite{ploschner2015seeing}. Various strategies for two-dimensional (2D) MMF imaging have been proposed utilizing wavefront shaping \cite{stibuurek2023110,leite2021observing}, speckle imaging \cite{choi2012scanner,caravaca2019hybrid}, deep learning \cite{rahmani2022learning,chen2023deep}, and compressive sensing \cite{amitonova2018compressive, amitonova2020endo}. Random speckle illumination provides a nearly ideal basis for compressive sensing approaches of imaging \cite{liutkus2014imaging}. Speckle patterns, such as those generated at the MMF output, can be used as structured illumination for super-resolution structural illumination microscopy (SIM) \cite{prakash2024resolution}. Random speckle illumination provides a resolution about two times better than that of conventional wide-field microscopy \cite{mudry2012structured}. Sparsity-based compressive sensing can further enhance the resolution beyond the physical limit of the microscope \cite{gazit_super-resolution_2009, sidorenko_sparsity-based_2015}. It has been recently shown that 2D compressive MMF imaging provides resolution beyond the diffraction limit using fewer measurements than the Nyquist limit and is applicable for different types of samples \cite{abrashitova2022high,lochocki2022epi}.


Transitioning from 2D to three-dimensional (3D) imaging is a captivating and impactful progression because 3D structure of an object contains far more information. 3D imaging through an MMF can be realized by using wavefront shaping: 3D raster scan at different axial planes \cite{wen2023single,loterie2015digital}, time-of-flight detection \cite{stellinga2021time}, and computational reconstruction \cite{dong2022modulated} have been demonstrated. However, all these methods rely on wavefront shaping and come with various drawbacks including complex setup and optimization procedures as well as low imaging speed. Other approaches to reconstructing 3D samples require the knowledge of the full transmission matrix of a fiber to decode depth information based on the propagation of light numerically \cite{lee2022confocal,sun2022quantitative}. These methods either require direct illumination to the sample, limiting the applicability for \emph{in vivo} imaging, or are only effective for coherent imaging, restricting the applicable types of samples to non-fluorescent.

Here, we present 3D imaging through an MMF by using a single 2D raster scan, random-speckle illumination, and single-pixel detection. Three-dimensional information is reconstructed using computational compressive sensing. A 3 times smaller FWHM of the point spread function along the axial direction than the diffraction limit is shown experimentally. We investigate the robustness of the proposed approach to volumetric light scattering effects and experimentally demonstrate 3D imaging of fluorescence beads in scattering media. The proposed approach is compatible with all kinds of MMF probes, including the flexible multicore-multimode fiber~\cite{lyu2023sub}. Therefore, our work provides an effective, simple, and robust 3D imaging method through an MMF paving the way for a flexible super-resolution high-speed endoscope.

\section{Methods}

\subsection{Experimental setup}

\begin{figure}[h!]
\centering
\includegraphics[width = 0.48\textwidth]{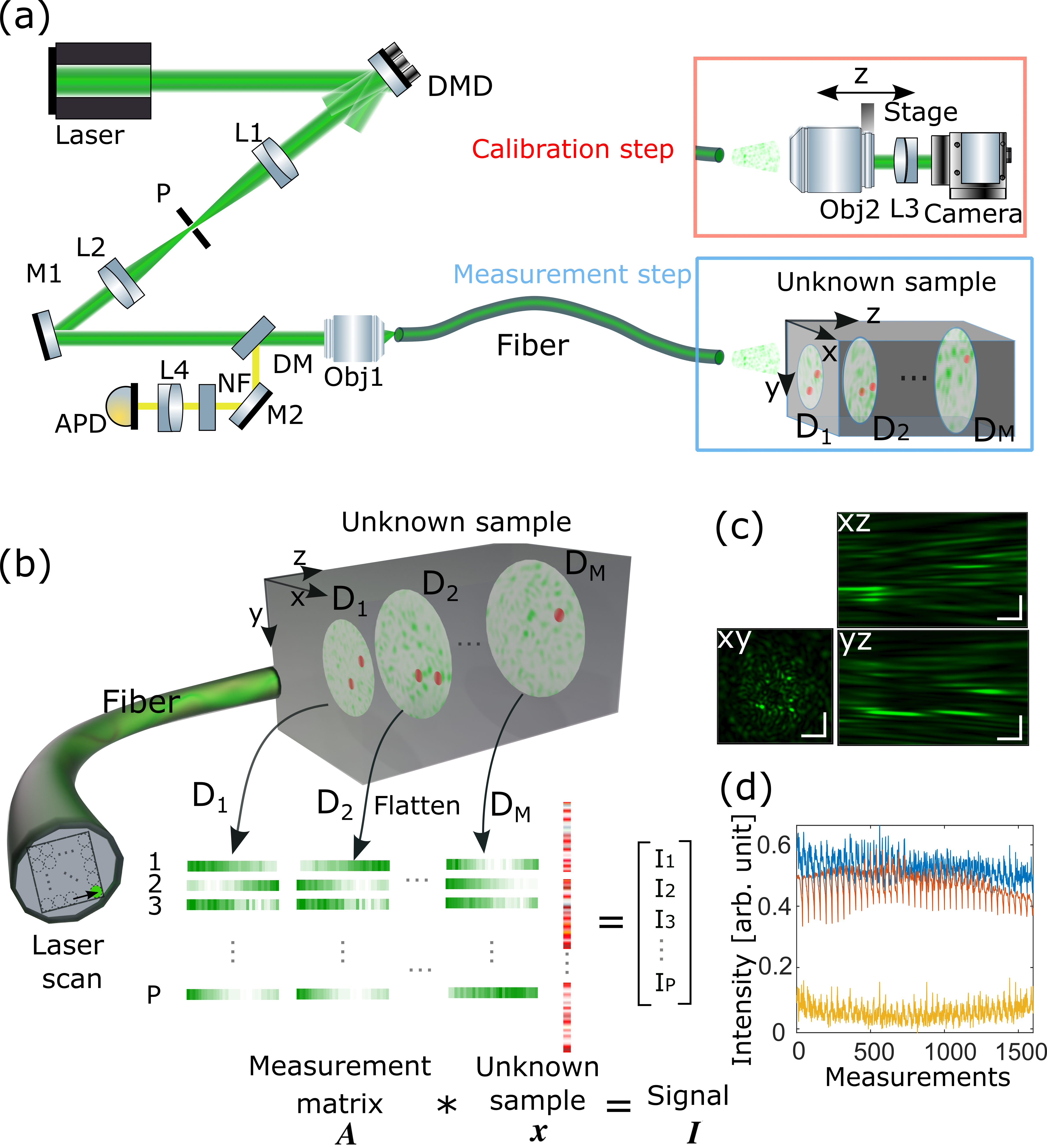}
\caption{\label{scheme}{ Illustration of the proposed approach of sub-diffraction 3D imaging through an MMF. (a)~Experimental setup. Light scans across the input facet of the MMF by the DMD. The speckles at various output planes $D_1 ...D_M$ are captured by the camera during the pre-calibration step as shown in the red box. The total fluorescence response $I_{p}$ is collected by the same fiber and registered by the APD. L, lenses; M, mirrors; DMD, digital micromirror device; Obj, objectives; P, pinhole; NF, notch filter; APD, avalanche photodiode. (b)~General idea and computational workflow. For each laser scan position $p$ at the input facet, a 3D speckle pattern illuminates the sample at planes $D_{1}$ to $D_{M}$. Those $P$ different 3D illumination patterns are flattened, connected head-to-tail, and stacked as a 2D measurement matrix $\boldsymbol{A}$. The unknown 3D sample is reshaped to vector $\boldsymbol{x}$. (c) The maximum intensity projections in $xy$-plane, $yz$-plane, and $xz$-plane of one 3D speckle pattern. The scale bars are 10 $\mu$m. (d) The signal $\boldsymbol{I}$ (yellow line) is calculated by substituting the measured intensity without the sample (red line) from the measured intensity with the sample (blue line). }}
\end{figure}

The experimental setup is shown in Fig.~\ref{scheme}(a). Experiments were performed on a round-core step-index MMF with a diameter of $50~\mu$m, and an NA of $0.22$ (Thorlabs). The fiber length is around 20 cm. For illumination, we use a continuous wave (CW) linear polarized Nd:YAG laser [Cobolt Samba] with a wavelength $\lambda = 532$~nm. The laser has a coherence length of 95 m which is longer than the maximum path length difference in the MMF to form fully developed speckle patterns on the output.
The digital micromirror device (DMD) from Texas Instrument driven by the DLP V-9501 VIS module (Vialux) is used to control the coupling to the MMF.
The DMD has a grid of 1920x1080 tilting micromirrors that can rotate individually to +12° (on state) or -12° (off state). By arranging these mirrors in either the on or off state, the DMD can function as a grating, with the grating's orientation determining the laser's position. The DMD pattern is imaged on the back focal plane of Objective1 (Olympus, 20x, NA = 0.4) by a 4f system consisting of lenses L1 (f1 = 150 mm) and L2 (f2 = 100 mm). The pinhole (P1) only allows the +1st diffraction order of the DMD pass. The Objective1 couples light to the MMF. Changing the grating direction, the laser beam can be scanned in a zigzag pattern across the fiber input facet. At the distal fiber facet, the speckles at different planes $D_1 ...D_M$ interact with the sample and excite the fluorescence response. \textcolor{Mycolor}{We use DMD for scanning because we need to be compatible with experiments based on WFS. One can choose other equipment with a faster response speed, such as an acousto-optical deflector or an electro-optical deflector.}

The experimental process is divided into two steps: the calibration step to record the speckles and the measurement step to detect the total excited fluorescence. In the calibration step, Objective2 (Leica, 63x, NA = 0.75) and lens L3 (f3 = 250 mm) image the fiber distal facet on the camera. By moving the motorized stage (Thorlabs Z912B) with Objective2 along the axial direction, different planes $D_1 ...D_M$ are imaged on the camera. Capture speckles for all $P$ scanning positions at the first imaging plane, and repeat this process for all $M$ imaging planes by moving Objective2. In the measurement step, an unknown sample is put right after the MMF. Light from the fiber excites the fluorescence. The total fluorescence response is collected by the same fiber, selected by the dichroic mirror (DM), and focused by lens L4 {(f4 = $60$~mm)} on an avalanche photodiode (Thorlabs APD440A). The notch filter (NF) blocks the pump light. The $P$ intensity values (shown in Fig.~\ref{scheme}(d)) are measured twice: once without the sample (red line), and then with the sample (blue line). Subtracting the results of two measurements yields the signal vector (yellow line). \textcolor{Mycolor}{The high background comes from various sources, such as fiber autofluorescence and Raman scattering.} In this and all following experiments, we kept using the $40\times40$ 2D scanning grid on the fiber input facet and recorded $1600$ intensities on the APD. In other words, the number of measurements $P = 1600$ is the same for all the experiments. The samples were additionally visualized by a conventional high-NA microscope to provide the reference images.

The expected diffraction-limited axial point spread function defined as FWHM criteria at the fiber output facet can be calculated for our fiber-based microscope as~\cite{goodman2007speckle,born2019principles}:
\begin{equation}
\label{axialdiffractionlimit}
D_\text{z}\approx 1.77\frac{\lambda }{\text{NA}^2}\approx 19.5~\mu\text{m}.
\end{equation}
The experimental characterization of the 3D illumination patterns for each position at the MMF input facet is achieved by recording intensity distributions on the fiber output with the motorized stage, the Objective2, and the camera. The intensity profiles of a 3D speckle pattern in a transverse $xy$-plane and axial $xz$- and $yz$-plane are shown in Fig.~\ref{scheme}(c). The scale bars are 10 $\mu$m. 
We experimentally measured the average size of the speckles generated by the MMF as presented and discussed in Supplementary~S1. Our experiments confirm that the speckle size in the axial ($z$-) direction is $21.2~\mu$m and in the lateral direction is $1.2~\mu\text{m}$, which are very close to the theoretical diffraction limit. 

\subsection{Computational reconstruction}
The main principle of 3D MMF computational imaging is shown in Fig.~\ref{scheme}(b). In contrast to conventional 3D raster scan imaging, the proposed approach does not require point-by-point illumination of the whole 3D region. We computationally reconstruct 3D image of a sample from the set of total fluorescence intensities collected for a single 2D raster scan performed at the MMF input facet. We use random patterns generated by an MMF and employ the fact that the speckle patterns decorrelate while propagating in free space: 2D speckle patterns appearing in different transverse planes on the MMF output are uncorrelated. 
Light scan on the input fiber facet (marked by the green dot in Fig.~\ref{scheme}(b)) gives rise to a complex 3D speckle pattern at the fiber output. Light interaction with the fluorescent sample can be described as if each 2D sample plane ($D_m$) is \textcolor{Mycolor}{multiplied} with the respective 2D speckle projection. The total fluorescence response, $I_{p}$, collected by the same MMF, is the linear combination of the contributions from all 2D planes $D_1 ... D_M$. 

For each scanning point at the input, named as a measurement, the speckles at planes $D_{1}$ to $D_{M}$ are recorded as $M$ matrices with $N \times N$ pixels and then flattened to $M$ rows of $N^2$ elements. These vectors are concatenated side-by-side to a single row of $1\times M\cdot N^2$ elements. The process is repeated $P$ times for $P$ scan points on the input facet leading to $P$ different 3D illumination patterns and consequently $P$ different rows.
By combining those $P$ rows vertically one below the other, one complete measurement matrix $\boldsymbol{A}$ ($P \times M\cdot N^2$) is built as shown in Fig.~\ref{scheme}(b). The unknown 3D sample is represented by one-dimensional vertical vector $\boldsymbol{x}$ ($M\cdot N^2 \times 1$) consisting of $M$ 2D sample planes (from $D_{1}$ to $D_{M}$), which are flattened and connected one below the other. Finally, the total fluorescence intensities $I_p$ recorded for each scanning position $p$ on the MMF input facet form vector $\boldsymbol{I}$. 
As a result, the measurements can be described as a simple matrix equation (also illustrated in Fig.~\ref{scheme}(b)):
\begin{equation}
\label{CS}
\boldsymbol{A}\boldsymbol{x} = \boldsymbol{I}.
\end{equation}
Solving this equation gives us $\boldsymbol{x}$, which is then split and reshaped back to $M$ matrices with $N \times N$ pixels resulting in a 3D image of the sample.
However, since we reconstruct a 3D structure from a single 2D scan, 
the number of measurements is much smaller than the number of sample pixels: $P \ll M\cdot N^2$. Therefore, the imaging problem converts to solving an underdetermined system of linear equations. The basis pursuit is generally used to solve the underdetermined problem by finding the most sparse solution with the minimum $\ell_1$-norm. We use the common SPGL1 solver which can be accessed and downloaded online \cite{spgl1site,BergFriedlander:2008}. It solves Eq. \ref{CS} by:
\begin{equation}
\label{spg}
\underset{\boldsymbol{x}}{\text{minimize}}\left \| \boldsymbol{x} \right \|_{1},\; s.t.~\boldsymbol{A}\boldsymbol{x} = \boldsymbol{I},
\end{equation}
where $\left \| \cdot \right \|_{1}$ is $\ell_1$-norm. Among the other compressive sensing algorithms, basis pursuit works best when the sample is sparse.
We used the central processing unit (CPU) of AMD Ryzen Threadripper 3960X, and Matlab 2022b for post-processing and image reconstruction.

\subsection{Sample preparation}\label{sample preparation}

We use fluorescently labeled polystyrene beads with a diameter of 1.14~$\mu$m to manipulate our 3D sample. To prepare the 3D fluorescence bead sample, we first affixed a coverslip by nail polish onto a metal plate with a central hole measuring 18 mm in diameter. Then, we prepared the 0.05$\%$ fluorescence stock by diluting the fluorescence bead with de-ionized water, 50$\%$ glycerol stock by diluting glycerol with de-ionized water, and 1$\%$ agarose stock by diluting agarose powder with de-ionized water. We vortexed the 50$\%$ glycerol stock for 1 minute to completely mix the glycerol with de-ionized water. Then, we mixed 100~$\mu$l 0.05$\%$ fluorescence stock, 50 $\mu$l 50$\%$ glycerol stock, and the corresponding concentration of scatter. The scatter is nanoparticles of Zinc oxide with < 40 nm average particle size. We added different volumes of scatter to adjust the final concentration of scatter (1:100, 1:250, 1:500, 1:2000, 1:10000) in the sample. The final concentration of agarose gel is 0.5$\%$. We vortex the mixed stock for 1 minute. Then, we add 150 1$\%$ agarose stock into the mixed stock and quickly pipette the mixed stock. This step needs to be finished within 10 seconds to avoid the agarose becoming solid. Afterwards, we added 200 $\mu$l mixed stock on the coverlip which is affixed on a metal plate. The glycerol in the mixed stock can prevent the agarose from drying out during measurement.

To measure the scattering coefficients of samples with different concentrations, we prepared the same mixed stock as above and added 1 ml of mixed stock to the cuvette. Then, the intensity before the sample and after the sample is measured, as described in Supplementary~S4. The scattering coefficients are calculated as $[0.18, 0.86, 1.98, 3.52, 7.52]$~cm$^{-1}$ according to the Lambert's law.

\section{Results}

\begin{figure}[t]
\centering
\includegraphics[width = 0.48\textwidth]{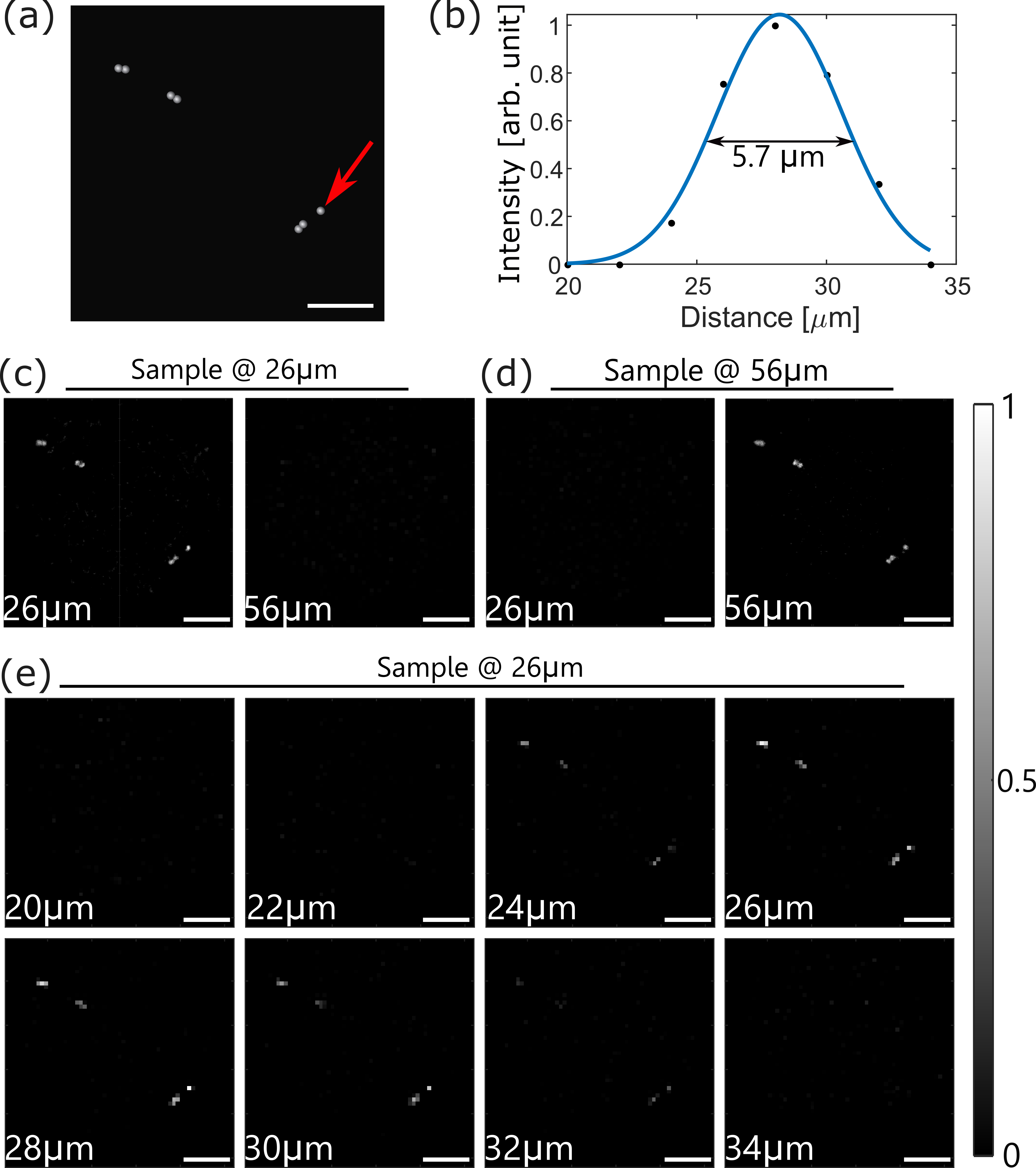}
\caption{\label{sample 2d + axial resolution}{Experimental results of the proposed MMF imaging and 3D localization approach for a 2D sample. (a) Rendered image of the sample based on high-resolution microscopic image. (b) Intensity of a single fluorescent bead (marked with red arrow in (a)) as a function of distance from the MMF output facet: experimental results (black dots) and Gaussian fit (blue line). The FWHM is 5.7~$\mu$m. (c-d) Imaging in two planes at a distance of 26~$\mu$m and 56~$\mu$m from the MMF facet if the 2D sample is at 26~$\mu$m plane (c) and 56~$\mu$m plane (d). (e) The imaging results in 8 planes from 20~$\mu$m to $34~\mu$m from the MMF facet. The sample is at 26~$\mu$m. The scale bars are 10~$\mu$m.}}
\end{figure}

In the first series of experiments, we demonstrate that the proposed approach allows for imaging a 2D fluorescent sample on the MMF facet and simultaneously precisely locating its axial position.
We use a single 2D sample consisting of fluorescent polystyrene particles deposited on a coverslip as presented in Fig.~\ref{sample 2d + axial resolution}(a). As a pre-calibration step, we record matrix $\boldsymbol{A}$ for $2$ planes separated by $30~\mu$m. During the measurements, we record the total fluorescent intensity collected by the MMF for every illumination pattern, $\boldsymbol{I}$. By solving Eq.~\ref{CS}, we reconstruct two images with $300\times300$ pixels that correspond to two separate planes along $z$-direction. We repeat the same experiment two times, first putting the sample in plane1 ($26~\mu$m from the fiber facet) and then in plane2 ($56~\mu$m from the fiber facet). The imaging results are presented in Fig.~\ref{sample 2d + axial resolution}(c) and ~\ref{sample 2d + axial resolution}(d), for the first and the second experiments, respectively. The data are normalized to the maximum detected intensity for both planes for each experiment separately. The 2D sample is visible only at the plane where it was actually placed during the experiment. In Fig.~~\ref{sample 2d + axial resolution}(c) the fluorescent beads are reconstructed in the first plane and the dark background appears in the second plane. The results are opposite if we put the sample in the second plan, as shown in Fig.~\ref{sample 2d + axial resolution}(d). We conclude that the proposed approach can image a 2D sample on the MMF output and simultaneously \textcolor{Mycolor}{differentiate the position of its depth}.

In the next set of experiments, we use the same 2D sample to characterize the axial resolution of the proposed 3D MMF compressive imaging approach. The 2D sample is put at $26~\mu$m distance from the MMF output facet. \textcolor{Mycolor}{The matrix $\boldsymbol{A}$ consists of sub-matrices connected head-to-tail, recorded at $8$ planes in $z$-direction from $20~\mu$m to $34~\mu$m with a step of $2~\mu$m in between. To decrease the compression rate, the 2D speckles are resized to $59 \times 59$ pixels.} The images are reconstructed at these planes at once using the single matrix $\boldsymbol{A}$. The results are shown in Fig.~\ref{sample 2d + axial resolution}(e).

The 2D sample is clearly imaged and located at the 26~$\mu$m plane where it is supposed to be. \textcolor{Mycolor}{We also see that the image is not precise only at 26~$\mu$m, but blurs forward and backward on the planes from 24 to 32~$\mu$m.} 
The more pronounced blurring in the forward direction can be explained by the uncertainty of the sample positioning due to the \textcolor{Mycolor}{up to $8~\mu$m backlash of the stage for moving the objective}. Correlation analysis of speckle patterns in different planes does not reveal a significant difference between forward and backward directions, as discussed in Supplementary~S3.
To quantify the effect, we plot the intensity of the brightest fluorescent bead (marked with the red arrow in Fig.~\ref{sample 2d + axial resolution}(a)) as a function of distance in $z$-direction. The results are presented in Fig.~\ref{sample 2d + axial resolution}(b) by black circles. A Gaussian fit (shown in blue) reveals a full width at half maximum (FWHM) of $5.7~\mu$m, which is $3$ times smaller than the axial diffraction limit of our MMF-based imaging system according to Eq.~\ref{axialdiffractionlimit}.


\textcolor{Mycolor}{ The experimentally measured the sub-diffraction axial width of the signal of a single bead suggests sub-diffraction resolution. To demonstrate that indeed two points can be differentiated, we performed a numerical experiment using experimentally measured 3D speckle patterns and an artificial sample. In our simulation, three objects with a size of $ 1 \times 1 \times 6 ~\mu$m$^3$ in $x$-, $y$-, and $z$-directions are positioned at the same lateral coordinates ($x$, $y$) and separated by $8$ and $12~\mu$m in the axial direction ($z$). The results are presented in Supplementary~S2. We see that the intensity at the midpoints between the peaks show a minimum and therefore all three objects are resolved. Our simulations demonstrate that the proposed imaging method can achieve super-resolution in the axial plane.}

\begin{figure*}[t]
\centering
\includegraphics[width = 0.7\textwidth]{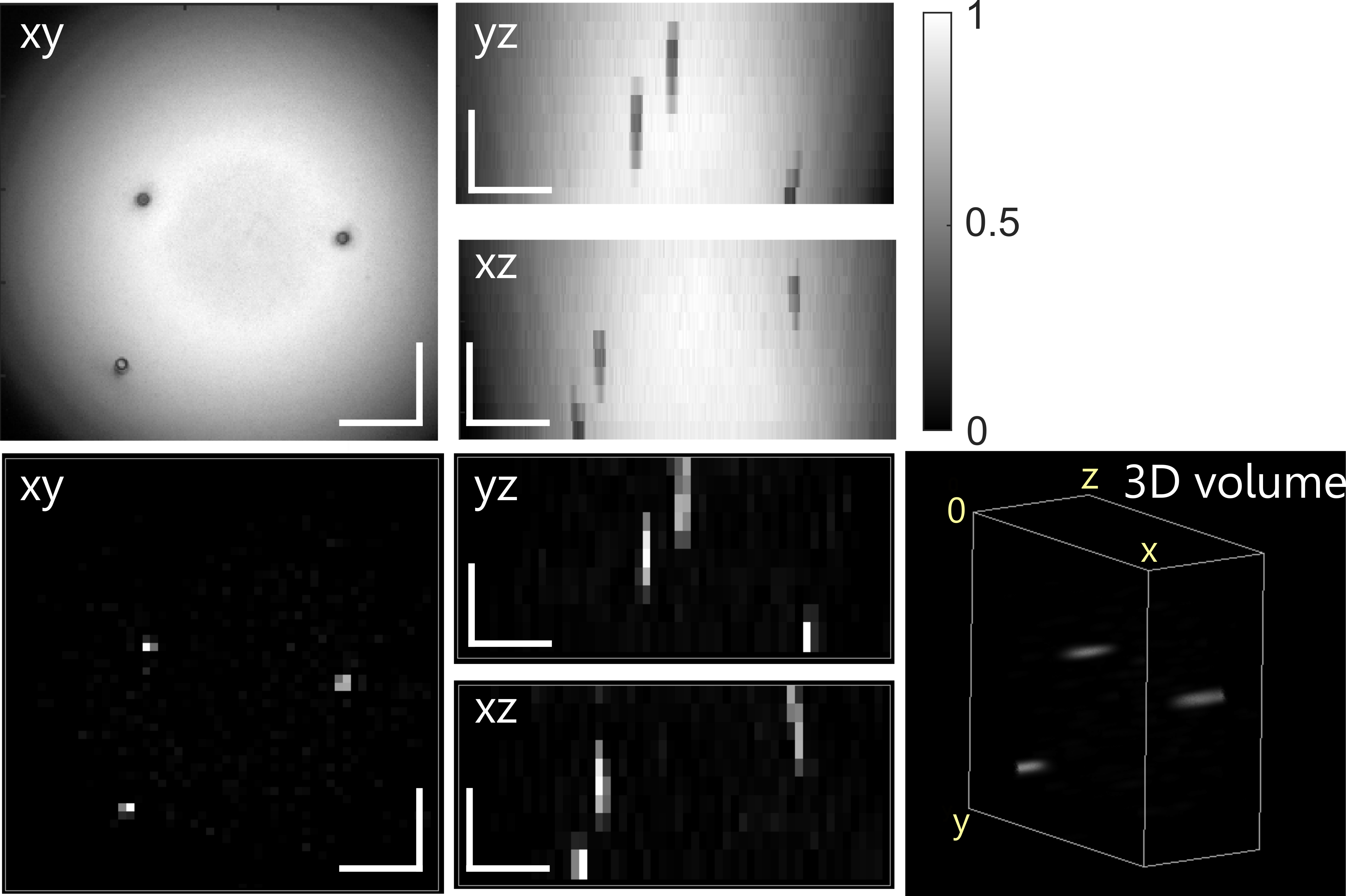}
\caption{\label{3d no scatter}{Experimental results of the proposed 3D imaging through an MMF approach for a 3D sample. \textcolor{Mycolor}{The minimum intensity projection of bright field images of the sample in $xy$-plane, $yz$-plane, and $xz$-plane are presented in the upper row.} The maximum intensity projections in $xy$-plane, $yz$-plane, $xz$-plane and the 3D volume \textcolor{Mycolor}{of the imaging results are shown in the blow}. The scale bars are 10 $\mu$m.}}
\end{figure*}

In the next series of experiments, we image a 3D sample consisting of three fluorescent beads randomly located at different positions in 3D. As a reference, the bright-field images of the sample are recorded with a high-NA microscope objective and LED illumination. The results are presented in the upper row of Fig.~\ref{3d no scatter}, where the minimal intensity projection in $xy$-plane, $yz$-plane, and $xz$-plane are shown.
For the proposed fiber-based 3D imaging, we use a single 2D scan on the MMF input and collect the total fluorescent signal. In these experiments, we reconstruct images at 11 planes with $2~\mu$m step in-between, starting at 52 $\mu$m from the fiber facet. The images are resized to $59\times59$ pixels. \textcolor{Mycolor}{Therefore, the sample is reconstructed within a field of view around $50 \times 50 \times 20~\mu$m$^3$ from $1600$ single-pixel measurements. The reconstruction results are presented in the lower row of Fig.~\ref{3d no scatter}. The 3D volumetric projection
is shown on the right and the maximum intensity projections in $xy$-plane, $yz$-plane, and $xz$-plane are shown on the left.} The three dots at different depths are reconstructed clearly with an estimated axial \textcolor{Mycolor}{width} of around $6~\mu$m which matches the previously calculated $5.7~\mu$m. The 3D volume size on the MMF output that was successfully imaged is $50\times\ 50\times 20~\mu$m$^3$.

In the final set of measurements, we experimentally investigate and demonstrate 3D imaging of scattering samples through the MMF by the proposed approach. \textcolor{Mycolor}{In biological imaging, one will encounter scattering, bringing two main problems. One is that the speckle illumination of the sample is inconsistent with the measured speckle. Another is that the fluorescence signal is reduced. Therefore, calibrating the behavior of our approach to scattering is important for biological application.} As a scatter, we use Zinc oxide nanoparticles added to the sample, as described in the Methods section. The scattering properties have been characterized in separate experiments as described in Supplementary~S4. 

\begin{figure*}[tb]
    
\centering
\includegraphics[width = 0.7\textwidth]{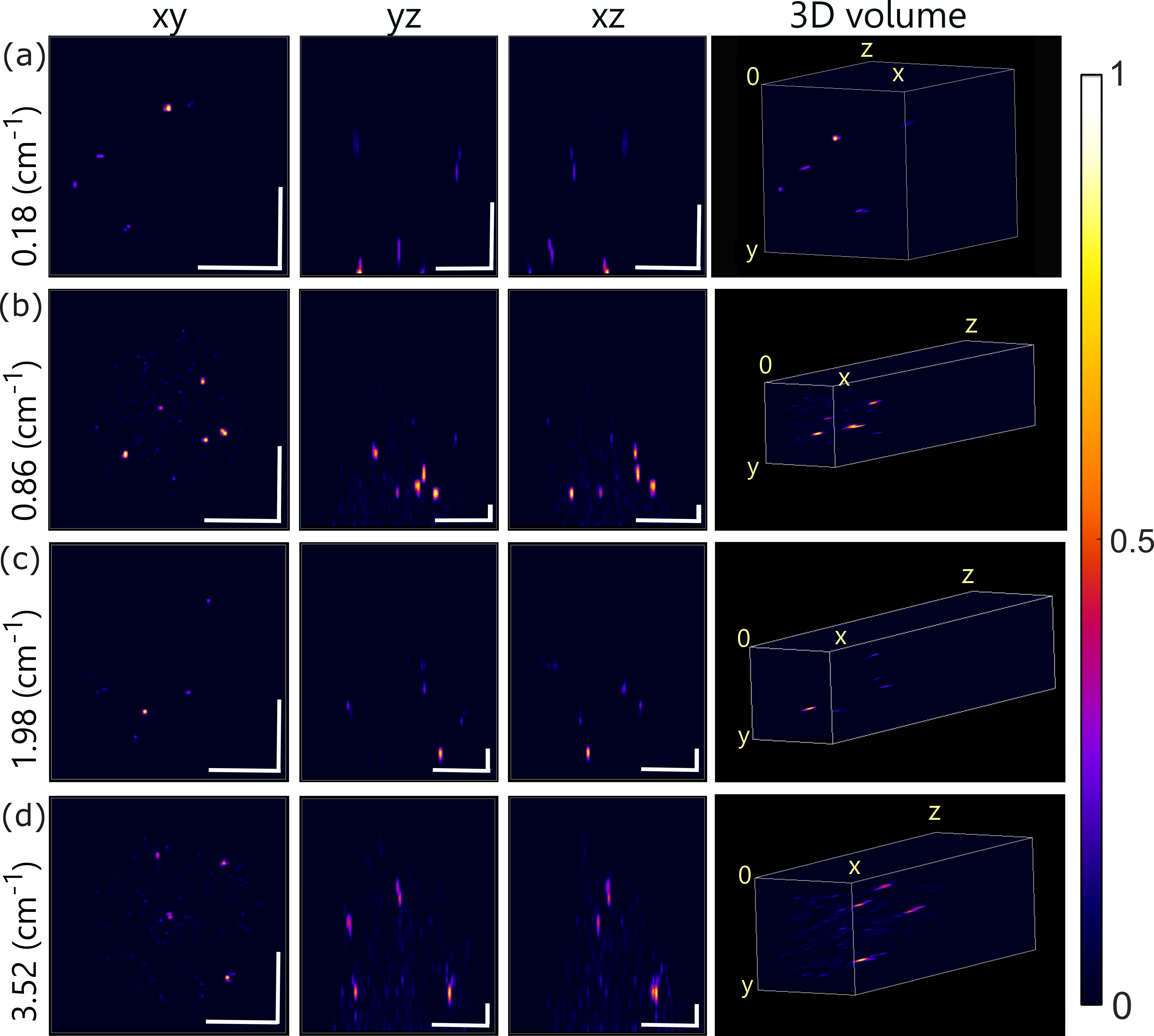}
\caption{\label{3d reconstruction}{(a-d) Experimental results of the proposed 3D imaging through an MMF approach for 3D scattering samples with scattering coefficients of $0.18$~cm$^{-1}$(a), $0.86$~cm$^{-1}$(b), $1.98$~cm$^{-1}$(c), $3.52$~cm$^{-1}$(d). The 3D volume projection, the maximum intensity projections in $xy$-plane, $yz$-plane, and $xz$-plane of the reconstruction are shown. The scale bars are 20 $\mu$m.}}
\end{figure*}
The experimental results of 3D imaging of scattering tissues through an MMF  are presented in Fig.~\ref{3d reconstruction}(a-d). Fluorescent beads are embedded into different samples with scattering coefficients of $[0.18, 0.86, 1.98, 3.52]$~cm$^{-1}$. The maximum intensity projections in $xy$-plane, $yz$-plane, and $xz$-plane are shown in the first, second, and third columns, respectively. The 3D volumes for each sample are presented in the fourth column. The scale bars are $20~\mu$m. The axial reconstruction range varies in the scattering samples due to the difficulty of manipulating with random 3D samples. As a result, the dimensions of the measurement matrix and the compression rate are different. Bright-field high-NA ground truth images (left) and images recorded through an MMF (right) at various planes in $z$-direction are shown in Supplementary Video1, 2, and 3, for samples with scattering coefficients of $[0.86, 1.98, 3.52]$~cm$^{-1}$, respectively. In bright-field reference images, the presence of a bead is indicated by darker colors. In contrast, in MMF-based imaging, the fluorescent signal from the bead is represented by whiter colors. Additionally, white circles indicate the locations of the beads.
Some representative planes where the beads are focused are also shown in Supplementary~S7. Comparing the ground truth and the images, we demonstrate the 3D reconstruction within a field of view around $50 \times 50 \times 140~\mu$m$^3$, based on $1600$ single-pixel measurements. 

\section{Discussion and conclusion}

We propose and experimentally demonstrate the new approach of 3D speckle-based compressive imaging through an MMF by a single 2D raster scan and a single-pixel detection. We explore random speckle decorrelation in free space propagation and use a 3D speckle matrix pre-recorded for various depths. The 3D image is reconstructed using computational compressive sensing. We show that the volumetric reconstruction overcomes the Abbe and Nyquist limits. A more than a 3-fold improvement of the FWHM of the point spread function along $z$-direction over the diffraction limit is shown experimentally. 

We compare the performance of the sparsity-based reconstruction algorithm with the conventional matrix pseudoinverse. The reconstruction results of the Moore-Penrose inverse algorithm on the same measurement matrix $\boldsymbol{A}$ and signal vector $\boldsymbol{I}$ are shown in Supplementary~S5. In contrast to $\ell_1$-norm minimization algorithm, the pseudoinverse does not incorporate the sparsity constraint leading to an image with a lower signal-to-noise ratio and the almost twice larger FWHM of the axial PSF of $10.3~\mu$m.

We investigate the performance of the proposed approach under noise by adding simulated noise to the measured signal $\boldsymbol{I}$. The imaging results of basis pursuit and pseudoinverse, with signal-to-noise ratio from 25 $\text{dB}$ to 5 $\text{dB}$ are shown in Supplementary~S6. The $\ell_1$-norm minimization algorithm shows higher noise tolerance by reconstructing the sample with a superior signal-to-background ratio at 5 $\text{dB}$ compared to the pseudoinverse at 25 $\text{dB}$.

\textcolor{Mycolor}{ Despite its high potential, the method has limitations. Careful consideration of the compression ratio is necessary, and in 3D imaging, illumination for distant planes may be obstructed by samples at the same lateral position in nearer planes. 
The maximum 3D volume is constrained by the fiber size in the lateral plane and the signal collection efficiency in the axial direction for non-scattering samples. 
For scattering samples, the maximum depth is reduced, as the speckles are affected by light scattering within the sample. 
} 

The speed of the proposed 3D MMF imaging is equal to the speed of a single 2D raster scan. This implies that the imaging process can be exceptionally fast. Larger 3D volumes require longer time for pre-calibration and post-processing, but this will not impact the measurement time needed for collecting intensity data. It takes about several minutes limited by the camera frame rate for pre-calibration and CPU for post-processing. However, as pre-calibration is done once before the measurements and post-processing can be done after all the series of measurements, they do not limit the high-speed 3D imaging abilities of the proposed approach. The fast imaging speed is attributed to the use of a compressive sensing algorithm. The scan size determines the number of rows $P$ of the measurement matrix in the compressive sensing algorithm, and the image volume is proportional to $M\cdot N^2$. Therefore, when the scan size is fixed and the imaging volume is increased, it will result in a higher compression rate. A higher compression rate imposes greater demands on sparsity. \textcolor{Mycolor}{The required total number of measurements is directly proportional to the sparsity.} 
The experimental imaging speed is limited by the 2D scanning rate of the DMD, the read-out speed and the sensitivity of the APD. In our experiments, for all 2D samples and 3D samples, the size of the 2D scan was $40 \times 40$ and the DMD repetition rate was set to $100$ Hz, resulting in a measurement time of about $0.5$ minute per image. With the maximum DMD speed of $23$~kHz and a single 2D scan size of $40\times40$, it will take only $70$~ms to generate a 3D image
, without any changes to the procedure or hardware. This corresponds to video-rate 3D imaging of about $14$ volumes per second. 

The proposed approach requires no intricate wavefront shaping, thereby simplifying the overall experimental process and the setup. It is compatible with all kinds of MMFs, including the flexible multicore-multimode fiber probe \cite{lyu2023sub}. To simplify the pre-calibration step, instead of recording speckles plane by plane we can record a full-field in the single plane on the fiber output and then calculate light propagation in 3D. 
Our work provides an effective, simple, and robust 3D imaging method through an MMF paving the way for a flexible 3D super-resolution high-speed endoscope.

\section*{Funding}
We acknowledge the Nederlandse Organisatie voor Wetenschappelijk Onderzoek (WISE).

\section*{Acknowledgments}
This work has been partially carried out within ARCNL, a public-private partnership between UvA, VU, NWO, and ASML, and was financed by ‘Toeslag voor Topconsortia voor Kennis en Innovatie (TKI)’ from the Dutch Ministry of Economic Affairs and Climate Policy. We thank Marco Seynen for his help in programming the data acquisition software.

\section*{Disclosures}
The authors declare no conflicts of interest.

\section*{Data Availability Statement}
 Data underlying the results presented in this paper are not publicly available at this time but may be obtained from the authors upon reasonable request.

\nocite{*}
\bibliography{my}

\begin{thebibliography}{35}%
\makeatletter
\providecommand \@ifxundefined [1]{%
 \@ifx{#1\undefined}
}%
\providecommand \@ifnum [1]{%
 \ifnum #1\expandafter \@firstoftwo
 \else \expandafter \@secondoftwo
 \fi
}%
\providecommand \@ifx [1]{%
 \ifx #1\expandafter \@firstoftwo
 \else \expandafter \@secondoftwo
 \fi
}%
\providecommand \natexlab [1]{#1}%
\providecommand \enquote  [1]{``#1''}%
\providecommand \bibnamefont  [1]{#1}%
\providecommand \bibfnamefont [1]{#1}%
\providecommand \citenamefont [1]{#1}%
\providecommand \href@noop [0]{\@secondoftwo}%
\providecommand \href [0]{\begingroup \@sanitize@url \@href}%
\providecommand \@href[1]{\@@startlink{#1}\@@href}%
\providecommand \@@href[1]{\endgroup#1\@@endlink}%
\providecommand \@sanitize@url [0]{\catcode `\\12\catcode `\$12\catcode `\&12\catcode `\#12\catcode `\^12\catcode `\_12\catcode `\%12\relax}%
\providecommand \@@startlink[1]{}%
\providecommand \@@endlink[0]{}%
\providecommand \url  [0]{\begingroup\@sanitize@url \@url }%
\providecommand \@url [1]{\endgroup\@href {#1}{\urlprefix }}%
\providecommand \urlprefix  [0]{URL }%
\providecommand \Eprint [0]{\href }%
\providecommand \doibase [0]{http://dx.doi.org/}%
\providecommand \selectlanguage [0]{\@gobble}%
\providecommand \bibinfo  [0]{\@secondoftwo}%
\providecommand \bibfield  [0]{\@secondoftwo}%
\providecommand \translation [1]{[#1]}%
\providecommand \BibitemOpen [0]{}%
\providecommand \bibitemStop [0]{}%
\providecommand \bibitemNoStop [0]{.\EOS\space}%
\providecommand \EOS [0]{\spacefactor3000\relax}%
\providecommand \BibitemShut  [1]{\csname bibitem#1\endcsname}%
\let\auto@bib@innerbib\@empty
\bibitem [{\citenamefont {Ntziachristos}, \citenamefont {Bremer},\ and\ \citenamefont {Weissleder}(2003)}]{ntziachristos2003fluorescence}%
  \BibitemOpen
  \bibfield  {author} {\bibinfo {author} {\bibfnamefont {V.}~\bibnamefont {Ntziachristos}}, \bibinfo {author} {\bibfnamefont {C.}~\bibnamefont {Bremer}}, \ and\ \bibinfo {author} {\bibfnamefont {R.}~\bibnamefont {Weissleder}},\ }\bibfield  {title} {\enquote {\bibinfo {title} {Fluorescence imaging with near-infrared light: new technological advances that enable in vivo molecular imaging},}\ }\href@noop {} {\bibfield  {journal} {\bibinfo  {journal} {European radiology}\ }\textbf {\bibinfo {volume} {13}},\ \bibinfo {pages} {195--208} (\bibinfo {year} {2003})}\BibitemShut {NoStop}%
\bibitem [{\citenamefont {Lauwerends}\ \emph {et~al.}(2021)\citenamefont {Lauwerends}, \citenamefont {van Driel}, \citenamefont {de~Jong}, \citenamefont {Hardillo}, \citenamefont {Koljenovic}, \citenamefont {Puppels}, \citenamefont {Mezzanotte}, \citenamefont {L{\"o}wik}, \citenamefont {Rosenthal}, \citenamefont {Vahrmeijer} \emph {et~al.}}]{lauwerends2021real}%
  \BibitemOpen
  \bibfield  {author} {\bibinfo {author} {\bibfnamefont {L.~J.}\ \bibnamefont {Lauwerends}}, \bibinfo {author} {\bibfnamefont {P.~B.}\ \bibnamefont {van Driel}}, \bibinfo {author} {\bibfnamefont {R.~J.~B.}\ \bibnamefont {de~Jong}}, \bibinfo {author} {\bibfnamefont {J.~A.}\ \bibnamefont {Hardillo}}, \bibinfo {author} {\bibfnamefont {S.}~\bibnamefont {Koljenovic}}, \bibinfo {author} {\bibfnamefont {G.}~\bibnamefont {Puppels}}, \bibinfo {author} {\bibfnamefont {L.}~\bibnamefont {Mezzanotte}}, \bibinfo {author} {\bibfnamefont {C.~W.}\ \bibnamefont {L{\"o}wik}}, \bibinfo {author} {\bibfnamefont {E.~L.}\ \bibnamefont {Rosenthal}}, \bibinfo {author} {\bibfnamefont {A.~L.}\ \bibnamefont {Vahrmeijer}},  \emph {et~al.},\ }\bibfield  {title} {\enquote {\bibinfo {title} {Real-time fluorescence imaging in intraoperative decision making for cancer surgery},}\ }\href@noop {} {\bibfield  {journal} {\bibinfo  {journal} {The Lancet Oncology}\ }\textbf {\bibinfo {volume} {22}},\ \bibinfo {pages} {e186--e195} (\bibinfo {year}
  {2021})}\BibitemShut {NoStop}%
\bibitem [{\citenamefont {Galbraith}\ and\ \citenamefont {Galbraith}(2011)}]{galbraith2011super}%
  \BibitemOpen
  \bibfield  {author} {\bibinfo {author} {\bibfnamefont {C.~G.}\ \bibnamefont {Galbraith}}\ and\ \bibinfo {author} {\bibfnamefont {J.~A.}\ \bibnamefont {Galbraith}},\ }\bibfield  {title} {\enquote {\bibinfo {title} {Super-resolution microscopy at a glance},}\ }\href@noop {} {\bibfield  {journal} {\bibinfo  {journal} {Journal of cell science}\ }\textbf {\bibinfo {volume} {124}},\ \bibinfo {pages} {1607--1611} (\bibinfo {year} {2011})}\BibitemShut {NoStop}%
\bibitem [{\citenamefont {Hell}\ \emph {et~al.}(2015)\citenamefont {Hell}, \citenamefont {Sahl}, \citenamefont {Bates}, \citenamefont {Zhuang}, \citenamefont {Heintzmann}, \citenamefont {Booth}, \citenamefont {Bewersdorf}, \citenamefont {Shtengel}, \citenamefont {Hess}, \citenamefont {Tinnefeld} \emph {et~al.}}]{hell20152015}%
  \BibitemOpen
  \bibfield  {author} {\bibinfo {author} {\bibfnamefont {S.~W.}\ \bibnamefont {Hell}}, \bibinfo {author} {\bibfnamefont {S.~J.}\ \bibnamefont {Sahl}}, \bibinfo {author} {\bibfnamefont {M.}~\bibnamefont {Bates}}, \bibinfo {author} {\bibfnamefont {X.}~\bibnamefont {Zhuang}}, \bibinfo {author} {\bibfnamefont {R.}~\bibnamefont {Heintzmann}}, \bibinfo {author} {\bibfnamefont {M.~J.}\ \bibnamefont {Booth}}, \bibinfo {author} {\bibfnamefont {J.}~\bibnamefont {Bewersdorf}}, \bibinfo {author} {\bibfnamefont {G.}~\bibnamefont {Shtengel}}, \bibinfo {author} {\bibfnamefont {H.}~\bibnamefont {Hess}}, \bibinfo {author} {\bibfnamefont {P.}~\bibnamefont {Tinnefeld}},  \emph {et~al.},\ }\bibfield  {title} {\enquote {\bibinfo {title} {The 2015 super-resolution microscopy roadmap},}\ }\href@noop {} {\bibfield  {journal} {\bibinfo  {journal} {Journal of Physics D: Applied Physics}\ }\textbf {\bibinfo {volume} {48}},\ \bibinfo {pages} {443001} (\bibinfo {year} {2015})}\BibitemShut {NoStop}%
\bibitem [{\citenamefont {Urban}\ \emph {et~al.}(2011)\citenamefont {Urban}, \citenamefont {Willig}, \citenamefont {Hell},\ and\ \citenamefont {N{\"a}gerl}}]{urban2011sted}%
  \BibitemOpen
  \bibfield  {author} {\bibinfo {author} {\bibfnamefont {N.~T.}\ \bibnamefont {Urban}}, \bibinfo {author} {\bibfnamefont {K.~I.}\ \bibnamefont {Willig}}, \bibinfo {author} {\bibfnamefont {S.~W.}\ \bibnamefont {Hell}}, \ and\ \bibinfo {author} {\bibfnamefont {U.~V.}\ \bibnamefont {N{\"a}gerl}},\ }\bibfield  {title} {\enquote {\bibinfo {title} {Sted nanoscopy of actin dynamics in synapses deep inside living brain slices},}\ }\href@noop {} {\bibfield  {journal} {\bibinfo  {journal} {Biophysical journal}\ }\textbf {\bibinfo {volume} {101}},\ \bibinfo {pages} {1277--1284} (\bibinfo {year} {2011})}\BibitemShut {NoStop}%
\bibitem [{\citenamefont {Jing}\ \emph {et~al.}(2021)\citenamefont {Jing}, \citenamefont {Zhang}, \citenamefont {Yu}, \citenamefont {Lin},\ and\ \citenamefont {Qu}}]{jing2021super}%
  \BibitemOpen
  \bibfield  {author} {\bibinfo {author} {\bibfnamefont {Y.}~\bibnamefont {Jing}}, \bibinfo {author} {\bibfnamefont {C.}~\bibnamefont {Zhang}}, \bibinfo {author} {\bibfnamefont {B.}~\bibnamefont {Yu}}, \bibinfo {author} {\bibfnamefont {D.}~\bibnamefont {Lin}}, \ and\ \bibinfo {author} {\bibfnamefont {J.}~\bibnamefont {Qu}},\ }\bibfield  {title} {\enquote {\bibinfo {title} {Super-resolution microscopy: shedding new light on in vivo imaging},}\ }\href@noop {} {\bibfield  {journal} {\bibinfo  {journal} {Frontiers in Chemistry}\ }\textbf {\bibinfo {volume} {9}},\ \bibinfo {pages} {746900} (\bibinfo {year} {2021})}\BibitemShut {NoStop}%
\bibitem [{\citenamefont {Andresen}\ \emph {et~al.}(2016)\citenamefont {Andresen}, \citenamefont {Sivankutty}, \citenamefont {Tsvirkun}, \citenamefont {Bouwmans},\ and\ \citenamefont {Rigneault}}]{andresen2016ultrathin}%
  \BibitemOpen
  \bibfield  {author} {\bibinfo {author} {\bibfnamefont {E.~R.}\ \bibnamefont {Andresen}}, \bibinfo {author} {\bibfnamefont {S.}~\bibnamefont {Sivankutty}}, \bibinfo {author} {\bibfnamefont {V.}~\bibnamefont {Tsvirkun}}, \bibinfo {author} {\bibfnamefont {G.}~\bibnamefont {Bouwmans}}, \ and\ \bibinfo {author} {\bibfnamefont {H.}~\bibnamefont {Rigneault}},\ }\bibfield  {title} {\enquote {\bibinfo {title} {Ultrathin endoscopes based on multicore fibers and adaptive optics: a status review and perspectives},}\ }\href@noop {} {\bibfield  {journal} {\bibinfo  {journal} {Journal of biomedical optics}\ }\textbf {\bibinfo {volume} {21}},\ \bibinfo {pages} {121506--121506} (\bibinfo {year} {2016})}\BibitemShut {NoStop}%
\bibitem [{\citenamefont {Badt}\ and\ \citenamefont {Katz}(2022)}]{badt2022real}%
  \BibitemOpen
  \bibfield  {author} {\bibinfo {author} {\bibfnamefont {N.}~\bibnamefont {Badt}}\ and\ \bibinfo {author} {\bibfnamefont {O.}~\bibnamefont {Katz}},\ }\bibfield  {title} {\enquote {\bibinfo {title} {Real-time holographic lensless micro-endoscopy through flexible fibers via fiber bundle distal holography},}\ }\href@noop {} {\bibfield  {journal} {\bibinfo  {journal} {Nature Communications}\ }\textbf {\bibinfo {volume} {13}},\ \bibinfo {pages} {6055} (\bibinfo {year} {2022})}\BibitemShut {NoStop}%
\bibitem [{\citenamefont {Pl{\"o}schner}, \citenamefont {Tyc},\ and\ \citenamefont {{\v{C}}i{\v{z}}m{\'a}r}(2015)}]{ploschner2015seeing}%
  \BibitemOpen
  \bibfield  {author} {\bibinfo {author} {\bibfnamefont {M.}~\bibnamefont {Pl{\"o}schner}}, \bibinfo {author} {\bibfnamefont {T.}~\bibnamefont {Tyc}}, \ and\ \bibinfo {author} {\bibfnamefont {T.}~\bibnamefont {{\v{C}}i{\v{z}}m{\'a}r}},\ }\bibfield  {title} {\enquote {\bibinfo {title} {Seeing through chaos in multimode fibres},}\ }\href@noop {} {\bibfield  {journal} {\bibinfo  {journal} {Nature Photonics}\ }\textbf {\bibinfo {volume} {9}},\ \bibinfo {pages} {529--535} (\bibinfo {year} {2015})}\BibitemShut {NoStop}%
\bibitem [{\citenamefont {Stib{\r u}rek}\ \emph {et~al.}(2023)\citenamefont {Stib{\r u}rek}, \citenamefont {Ondr{\'a}{\v{c}}kov{\'a}}, \citenamefont {Tu{\v{c}}kov{\'a}}, \citenamefont {Turtaev}, \citenamefont {{\v{S}}iler}, \citenamefont {Pik{\'a}lek}, \citenamefont {J{\'a}kl}, \citenamefont {Gomes}, \citenamefont {Krej{\v{c}}{\'\i}}, \citenamefont {Kolb{\'a}bkov{\'a}} \emph {et~al.}}]{stibuurek2023110}%
  \BibitemOpen
  \bibfield  {author} {\bibinfo {author} {\bibfnamefont {M.}~\bibnamefont {Stib{\r u}rek}}, \bibinfo {author} {\bibfnamefont {P.}~\bibnamefont {Ondr{\'a}{\v{c}}kov{\'a}}}, \bibinfo {author} {\bibfnamefont {T.}~\bibnamefont {Tu{\v{c}}kov{\'a}}}, \bibinfo {author} {\bibfnamefont {S.}~\bibnamefont {Turtaev}}, \bibinfo {author} {\bibfnamefont {M.}~\bibnamefont {{\v{S}}iler}}, \bibinfo {author} {\bibfnamefont {T.}~\bibnamefont {Pik{\'a}lek}}, \bibinfo {author} {\bibfnamefont {P.}~\bibnamefont {J{\'a}kl}}, \bibinfo {author} {\bibfnamefont {A.}~\bibnamefont {Gomes}}, \bibinfo {author} {\bibfnamefont {J.}~\bibnamefont {Krej{\v{c}}{\'\i}}}, \bibinfo {author} {\bibfnamefont {P.}~\bibnamefont {Kolb{\'a}bkov{\'a}}},  \emph {et~al.},\ }\bibfield  {title} {\enquote {\bibinfo {title} {110 $\mu$m thin endo-microscope for deep-brain in vivo observations of neuronal connectivity, activity and blood flow dynamics},}\ }\href@noop {} {\bibfield  {journal} {\bibinfo  {journal} {Nature Communications}\ }\textbf {\bibinfo {volume}
  {14}},\ \bibinfo {pages} {1897} (\bibinfo {year} {2023})}\BibitemShut {NoStop}%
\bibitem [{\citenamefont {Leite}\ \emph {et~al.}(2021)\citenamefont {Leite}, \citenamefont {Turtaev}, \citenamefont {Boonzajer~Flaes},\ and\ \citenamefont {{\v{C}}i{\v{z}}m{\'a}r}}]{leite2021observing}%
  \BibitemOpen
  \bibfield  {author} {\bibinfo {author} {\bibfnamefont {I.~T.}\ \bibnamefont {Leite}}, \bibinfo {author} {\bibfnamefont {S.}~\bibnamefont {Turtaev}}, \bibinfo {author} {\bibfnamefont {D.~E.}\ \bibnamefont {Boonzajer~Flaes}}, \ and\ \bibinfo {author} {\bibfnamefont {T.}~\bibnamefont {{\v{C}}i{\v{z}}m{\'a}r}},\ }\bibfield  {title} {\enquote {\bibinfo {title} {Observing distant objects with a multimode fiber-based holographic endoscope},}\ }\href@noop {} {\bibfield  {journal} {\bibinfo  {journal} {APL Photonics}\ }\textbf {\bibinfo {volume} {6}},\ \bibinfo {pages} {036112} (\bibinfo {year} {2021})}\BibitemShut {NoStop}%
\bibitem [{\citenamefont {Choi}\ \emph {et~al.}(2012)\citenamefont {Choi}, \citenamefont {Yoon}, \citenamefont {Kim}, \citenamefont {Yang}, \citenamefont {Fang-Yen}, \citenamefont {Dasari}, \citenamefont {Lee},\ and\ \citenamefont {Choi}}]{choi2012scanner}%
  \BibitemOpen
  \bibfield  {author} {\bibinfo {author} {\bibfnamefont {Y.}~\bibnamefont {Choi}}, \bibinfo {author} {\bibfnamefont {C.}~\bibnamefont {Yoon}}, \bibinfo {author} {\bibfnamefont {M.}~\bibnamefont {Kim}}, \bibinfo {author} {\bibfnamefont {T.~D.}\ \bibnamefont {Yang}}, \bibinfo {author} {\bibfnamefont {C.}~\bibnamefont {Fang-Yen}}, \bibinfo {author} {\bibfnamefont {R.~R.}\ \bibnamefont {Dasari}}, \bibinfo {author} {\bibfnamefont {K.~J.}\ \bibnamefont {Lee}}, \ and\ \bibinfo {author} {\bibfnamefont {W.}~\bibnamefont {Choi}},\ }\bibfield  {title} {\enquote {\bibinfo {title} {Scanner-free and wide-field endoscopic imaging by using a single multimode optical fiber},}\ }\href@noop {} {\bibfield  {journal} {\bibinfo  {journal} {Physical review letters}\ }\textbf {\bibinfo {volume} {109}},\ \bibinfo {pages} {203901} (\bibinfo {year} {2012})}\BibitemShut {NoStop}%
\bibitem [{\citenamefont {Caravaca-Aguirre}\ \emph {et~al.}(2019)\citenamefont {Caravaca-Aguirre}, \citenamefont {Singh}, \citenamefont {Labouesse}, \citenamefont {Baratta}, \citenamefont {Piestun},\ and\ \citenamefont {Bossy}}]{caravaca2019hybrid}%
  \BibitemOpen
  \bibfield  {author} {\bibinfo {author} {\bibfnamefont {A.~M.}\ \bibnamefont {Caravaca-Aguirre}}, \bibinfo {author} {\bibfnamefont {S.}~\bibnamefont {Singh}}, \bibinfo {author} {\bibfnamefont {S.}~\bibnamefont {Labouesse}}, \bibinfo {author} {\bibfnamefont {M.~V.}\ \bibnamefont {Baratta}}, \bibinfo {author} {\bibfnamefont {R.}~\bibnamefont {Piestun}}, \ and\ \bibinfo {author} {\bibfnamefont {E.}~\bibnamefont {Bossy}},\ }\bibfield  {title} {\enquote {\bibinfo {title} {Hybrid photoacoustic-fluorescence microendoscopy through a multimode fiber using speckle illumination},}\ }\href@noop {} {\bibfield  {journal} {\bibinfo  {journal} {Apl Photonics}\ }\textbf {\bibinfo {volume} {4}} (\bibinfo {year} {2019})}\BibitemShut {NoStop}%
\bibitem [{\citenamefont {Rahmani}\ \emph {et~al.}(2022)\citenamefont {Rahmani}, \citenamefont {Oguz}, \citenamefont {Tegin}, \citenamefont {Hsieh}, \citenamefont {Psaltis},\ and\ \citenamefont {Moser}}]{rahmani2022learning}%
  \BibitemOpen
  \bibfield  {author} {\bibinfo {author} {\bibfnamefont {B.}~\bibnamefont {Rahmani}}, \bibinfo {author} {\bibfnamefont {I.}~\bibnamefont {Oguz}}, \bibinfo {author} {\bibfnamefont {U.}~\bibnamefont {Tegin}}, \bibinfo {author} {\bibfnamefont {J.-l.}\ \bibnamefont {Hsieh}}, \bibinfo {author} {\bibfnamefont {D.}~\bibnamefont {Psaltis}}, \ and\ \bibinfo {author} {\bibfnamefont {C.}~\bibnamefont {Moser}},\ }\bibfield  {title} {\enquote {\bibinfo {title} {Learning to image and compute with multimode optical fibers},}\ }\href@noop {} {\bibfield  {journal} {\bibinfo  {journal} {Nanophotonics}\ }\textbf {\bibinfo {volume} {11}},\ \bibinfo {pages} {1071--1082} (\bibinfo {year} {2022})}\BibitemShut {NoStop}%
\bibitem [{\citenamefont {Chen}\ \emph {et~al.}(2023)\citenamefont {Chen}, \citenamefont {Song}, \citenamefont {Wu}, \citenamefont {Lin},\ and\ \citenamefont {Huang}}]{chen2023deep}%
  \BibitemOpen
  \bibfield  {author} {\bibinfo {author} {\bibfnamefont {Y.}~\bibnamefont {Chen}}, \bibinfo {author} {\bibfnamefont {B.}~\bibnamefont {Song}}, \bibinfo {author} {\bibfnamefont {J.}~\bibnamefont {Wu}}, \bibinfo {author} {\bibfnamefont {W.}~\bibnamefont {Lin}}, \ and\ \bibinfo {author} {\bibfnamefont {W.}~\bibnamefont {Huang}},\ }\bibfield  {title} {\enquote {\bibinfo {title} {Deep learning for efficiently imaging through the localized speckle field of a multimode fiber},}\ }\href@noop {} {\bibfield  {journal} {\bibinfo  {journal} {Applied Optics}\ }\textbf {\bibinfo {volume} {62}},\ \bibinfo {pages} {266--274} (\bibinfo {year} {2023})}\BibitemShut {NoStop}%
\bibitem [{\citenamefont {Amitonova}\ and\ \citenamefont {De~Boer}(2018)}]{amitonova2018compressive}%
  \BibitemOpen
  \bibfield  {author} {\bibinfo {author} {\bibfnamefont {L.~V.}\ \bibnamefont {Amitonova}}\ and\ \bibinfo {author} {\bibfnamefont {J.~F.}\ \bibnamefont {De~Boer}},\ }\bibfield  {title} {\enquote {\bibinfo {title} {Compressive imaging through a multimode fiber},}\ }\href@noop {} {\bibfield  {journal} {\bibinfo  {journal} {Optics letters}\ }\textbf {\bibinfo {volume} {43}},\ \bibinfo {pages} {5427--5430} (\bibinfo {year} {2018})}\BibitemShut {NoStop}%
\bibitem [{\citenamefont {Amitonova}\ and\ \citenamefont {de~Boer}(2020)}]{amitonova2020endo}%
  \BibitemOpen
  \bibfield  {author} {\bibinfo {author} {\bibfnamefont {L.~V.}\ \bibnamefont {Amitonova}}\ and\ \bibinfo {author} {\bibfnamefont {J.~F.}\ \bibnamefont {de~Boer}},\ }\bibfield  {title} {\enquote {\bibinfo {title} {Endo-microscopy beyond the abbe and nyquist limits},}\ }\href@noop {} {\bibfield  {journal} {\bibinfo  {journal} {Light: Science \& Applications}\ }\textbf {\bibinfo {volume} {9}},\ \bibinfo {pages} {81} (\bibinfo {year} {2020})}\BibitemShut {NoStop}%
\bibitem [{\citenamefont {Liutkus}\ \emph {et~al.}(2014)\citenamefont {Liutkus}, \citenamefont {Martina}, \citenamefont {Popoff}, \citenamefont {Chardon}, \citenamefont {Katz}, \citenamefont {Lerosey}, \citenamefont {Gigan}, \citenamefont {Daudet},\ and\ \citenamefont {Carron}}]{liutkus2014imaging}%
  \BibitemOpen
  \bibfield  {author} {\bibinfo {author} {\bibfnamefont {A.}~\bibnamefont {Liutkus}}, \bibinfo {author} {\bibfnamefont {D.}~\bibnamefont {Martina}}, \bibinfo {author} {\bibfnamefont {S.}~\bibnamefont {Popoff}}, \bibinfo {author} {\bibfnamefont {G.}~\bibnamefont {Chardon}}, \bibinfo {author} {\bibfnamefont {O.}~\bibnamefont {Katz}}, \bibinfo {author} {\bibfnamefont {G.}~\bibnamefont {Lerosey}}, \bibinfo {author} {\bibfnamefont {S.}~\bibnamefont {Gigan}}, \bibinfo {author} {\bibfnamefont {L.}~\bibnamefont {Daudet}}, \ and\ \bibinfo {author} {\bibfnamefont {I.}~\bibnamefont {Carron}},\ }\bibfield  {title} {\enquote {\bibinfo {title} {Imaging with nature: Compressive imaging using a multiply scattering medium},}\ }\href@noop {} {\bibfield  {journal} {\bibinfo  {journal} {Scientific reports}\ }\textbf {\bibinfo {volume} {4}},\ \bibinfo {pages} {5552} (\bibinfo {year} {2014})}\BibitemShut {NoStop}%
\bibitem [{\citenamefont {Prakash}\ \emph {et~al.}(2024)\citenamefont {Prakash}, \citenamefont {Baddeley}, \citenamefont {Eggeling}, \citenamefont {Fiolka}, \citenamefont {Heintzmann}, \citenamefont {Manley}, \citenamefont {Radenovic}, \citenamefont {Smith}, \citenamefont {Shroff},\ and\ \citenamefont {Schermelleh}}]{prakash2024resolution}%
  \BibitemOpen
  \bibfield  {author} {\bibinfo {author} {\bibfnamefont {K.}~\bibnamefont {Prakash}}, \bibinfo {author} {\bibfnamefont {D.}~\bibnamefont {Baddeley}}, \bibinfo {author} {\bibfnamefont {C.}~\bibnamefont {Eggeling}}, \bibinfo {author} {\bibfnamefont {R.}~\bibnamefont {Fiolka}}, \bibinfo {author} {\bibfnamefont {R.}~\bibnamefont {Heintzmann}}, \bibinfo {author} {\bibfnamefont {S.}~\bibnamefont {Manley}}, \bibinfo {author} {\bibfnamefont {A.}~\bibnamefont {Radenovic}}, \bibinfo {author} {\bibfnamefont {C.}~\bibnamefont {Smith}}, \bibinfo {author} {\bibfnamefont {H.}~\bibnamefont {Shroff}}, \ and\ \bibinfo {author} {\bibfnamefont {L.}~\bibnamefont {Schermelleh}},\ }\bibfield  {title} {\enquote {\bibinfo {title} {Resolution in super-resolution microscopy—definition, trade-offs and perspectives},}\ }\href@noop {} {\bibfield  {journal} {\bibinfo  {journal} {Nature Reviews Molecular Cell Biology}\ ,\ \bibinfo {pages} {1--6}} (\bibinfo {year} {2024})}\BibitemShut {NoStop}%
\bibitem [{\citenamefont {Mudry}\ \emph {et~al.}(2012)\citenamefont {Mudry}, \citenamefont {Belkebir}, \citenamefont {Girard}, \citenamefont {Savatier}, \citenamefont {Le~Moal}, \citenamefont {Nicoletti}, \citenamefont {Allain},\ and\ \citenamefont {Sentenac}}]{mudry2012structured}%
  \BibitemOpen
  \bibfield  {author} {\bibinfo {author} {\bibfnamefont {E.}~\bibnamefont {Mudry}}, \bibinfo {author} {\bibfnamefont {K.}~\bibnamefont {Belkebir}}, \bibinfo {author} {\bibfnamefont {J.}~\bibnamefont {Girard}}, \bibinfo {author} {\bibfnamefont {J.}~\bibnamefont {Savatier}}, \bibinfo {author} {\bibfnamefont {E.}~\bibnamefont {Le~Moal}}, \bibinfo {author} {\bibfnamefont {C.}~\bibnamefont {Nicoletti}}, \bibinfo {author} {\bibfnamefont {M.}~\bibnamefont {Allain}}, \ and\ \bibinfo {author} {\bibfnamefont {A.}~\bibnamefont {Sentenac}},\ }\bibfield  {title} {\enquote {\bibinfo {title} {Structured illumination microscopy using unknown speckle patterns},}\ }\href@noop {} {\bibfield  {journal} {\bibinfo  {journal} {Nature Photonics}\ }\textbf {\bibinfo {volume} {6}},\ \bibinfo {pages} {312--315} (\bibinfo {year} {2012})}\BibitemShut {NoStop}%
\bibitem [{\citenamefont {Gazit}\ \emph {et~al.}(2009)\citenamefont {Gazit}, \citenamefont {Szameit}, \citenamefont {Eldar},\ and\ \citenamefont {Segev}}]{gazit_super-resolution_2009}%
  \BibitemOpen
  \bibfield  {author} {\bibinfo {author} {\bibfnamefont {S.}~\bibnamefont {Gazit}}, \bibinfo {author} {\bibfnamefont {A.}~\bibnamefont {Szameit}}, \bibinfo {author} {\bibfnamefont {Y.~C.}\ \bibnamefont {Eldar}}, \ and\ \bibinfo {author} {\bibfnamefont {M.}~\bibnamefont {Segev}},\ }\bibfield  {title} {\enquote {\bibinfo {title} {Super-resolution and reconstruction of sparse sub-wavelength images},}\ }\href@noop {} {\bibfield  {journal} {\bibinfo  {journal} {Optics Express}\ }\textbf {\bibinfo {volume} {17}},\ \bibinfo {pages} {23920--23946} (\bibinfo {year} {2009})}\BibitemShut {NoStop}%
\bibitem [{\citenamefont {Sidorenko}\ \emph {et~al.}(2015)\citenamefont {Sidorenko}, \citenamefont {Kfir}, \citenamefont {Shechtman}, \citenamefont {Fleischer}, \citenamefont {Eldar}, \citenamefont {Segev},\ and\ \citenamefont {Cohen}}]{sidorenko_sparsity-based_2015}%
  \BibitemOpen
  \bibfield  {author} {\bibinfo {author} {\bibfnamefont {P.}~\bibnamefont {Sidorenko}}, \bibinfo {author} {\bibfnamefont {O.}~\bibnamefont {Kfir}}, \bibinfo {author} {\bibfnamefont {Y.}~\bibnamefont {Shechtman}}, \bibinfo {author} {\bibfnamefont {A.}~\bibnamefont {Fleischer}}, \bibinfo {author} {\bibfnamefont {Y.~C.}\ \bibnamefont {Eldar}}, \bibinfo {author} {\bibfnamefont {M.}~\bibnamefont {Segev}}, \ and\ \bibinfo {author} {\bibfnamefont {O.}~\bibnamefont {Cohen}},\ }\bibfield  {title} {\enquote {\bibinfo {title} {Sparsity-based super-resolved coherent diffraction imaging of one-dimensional objects},}\ }\href@noop {} {\bibfield  {journal} {\bibinfo  {journal} {Nature Communications}\ }\textbf {\bibinfo {volume} {6}},\ \bibinfo {pages} {8209} (\bibinfo {year} {2015})},\ \bibinfo {note} {number: 1 Publisher: Nature Publishing Group}\BibitemShut {NoStop}%
\bibitem [{\citenamefont {Abrashitova}\ and\ \citenamefont {Amitonova}(2022)}]{abrashitova2022high}%
  \BibitemOpen
  \bibfield  {author} {\bibinfo {author} {\bibfnamefont {K.}~\bibnamefont {Abrashitova}}\ and\ \bibinfo {author} {\bibfnamefont {L.~V.}\ \bibnamefont {Amitonova}},\ }\bibfield  {title} {\enquote {\bibinfo {title} {High-speed label-free multimode-fiber-based compressive imaging beyond the diffraction limit},}\ }\href@noop {} {\bibfield  {journal} {\bibinfo  {journal} {Optics Express}\ }\textbf {\bibinfo {volume} {30}},\ \bibinfo {pages} {10456--10469} (\bibinfo {year} {2022})}\BibitemShut {NoStop}%
\bibitem [{\citenamefont {Lochocki}\ \emph {et~al.}(2022)\citenamefont {Lochocki}, \citenamefont {Verweg}, \citenamefont {Hoozemans}, \citenamefont {de~Boer},\ and\ \citenamefont {Amitonova}}]{lochocki2022epi}%
  \BibitemOpen
  \bibfield  {author} {\bibinfo {author} {\bibfnamefont {B.}~\bibnamefont {Lochocki}}, \bibinfo {author} {\bibfnamefont {M.~V.}\ \bibnamefont {Verweg}}, \bibinfo {author} {\bibfnamefont {J.~J.}\ \bibnamefont {Hoozemans}}, \bibinfo {author} {\bibfnamefont {J.~F.}\ \bibnamefont {de~Boer}}, \ and\ \bibinfo {author} {\bibfnamefont {L.~V.}\ \bibnamefont {Amitonova}},\ }\bibfield  {title} {\enquote {\bibinfo {title} {Epi-fluorescence imaging of the human brain through a multimode fiber},}\ }\href@noop {} {\bibfield  {journal} {\bibinfo  {journal} {APL Photonics}\ }\textbf {\bibinfo {volume} {7}},\ \bibinfo {pages} {071301} (\bibinfo {year} {2022})}\BibitemShut {NoStop}%
\bibitem [{\citenamefont {Wen}\ \emph {et~al.}(2023)\citenamefont {Wen}, \citenamefont {Dong}, \citenamefont {Deng}, \citenamefont {Pang}, \citenamefont {Kaminski}, \citenamefont {Xu}, \citenamefont {Yan}, \citenamefont {Wang}, \citenamefont {Liu}, \citenamefont {Tang} \emph {et~al.}}]{wen2023single}%
  \BibitemOpen
  \bibfield  {author} {\bibinfo {author} {\bibfnamefont {Z.}~\bibnamefont {Wen}}, \bibinfo {author} {\bibfnamefont {Z.}~\bibnamefont {Dong}}, \bibinfo {author} {\bibfnamefont {Q.}~\bibnamefont {Deng}}, \bibinfo {author} {\bibfnamefont {C.}~\bibnamefont {Pang}}, \bibinfo {author} {\bibfnamefont {C.~F.}\ \bibnamefont {Kaminski}}, \bibinfo {author} {\bibfnamefont {X.}~\bibnamefont {Xu}}, \bibinfo {author} {\bibfnamefont {H.}~\bibnamefont {Yan}}, \bibinfo {author} {\bibfnamefont {L.}~\bibnamefont {Wang}}, \bibinfo {author} {\bibfnamefont {S.}~\bibnamefont {Liu}}, \bibinfo {author} {\bibfnamefont {J.}~\bibnamefont {Tang}},  \emph {et~al.},\ }\bibfield  {title} {\enquote {\bibinfo {title} {Single multimode fibre for in vivo light-field-encoded endoscopic imaging},}\ }\href@noop {} {\bibfield  {journal} {\bibinfo  {journal} {Nature Photonics}\ ,\ \bibinfo {pages} {1--9}} (\bibinfo {year} {2023})}\BibitemShut {NoStop}%
\bibitem [{\citenamefont {Loterie}\ \emph {et~al.}(2015)\citenamefont {Loterie}, \citenamefont {Farahi}, \citenamefont {Papadopoulos}, \citenamefont {Goy}, \citenamefont {Psaltis},\ and\ \citenamefont {Moser}}]{loterie2015digital}%
  \BibitemOpen
  \bibfield  {author} {\bibinfo {author} {\bibfnamefont {D.}~\bibnamefont {Loterie}}, \bibinfo {author} {\bibfnamefont {S.}~\bibnamefont {Farahi}}, \bibinfo {author} {\bibfnamefont {I.}~\bibnamefont {Papadopoulos}}, \bibinfo {author} {\bibfnamefont {A.}~\bibnamefont {Goy}}, \bibinfo {author} {\bibfnamefont {D.}~\bibnamefont {Psaltis}}, \ and\ \bibinfo {author} {\bibfnamefont {C.}~\bibnamefont {Moser}},\ }\bibfield  {title} {\enquote {\bibinfo {title} {Digital confocal microscopy through a multimode fiber},}\ }\href@noop {} {\bibfield  {journal} {\bibinfo  {journal} {Optics express}\ }\textbf {\bibinfo {volume} {23}},\ \bibinfo {pages} {23845--23858} (\bibinfo {year} {2015})}\BibitemShut {NoStop}%
\bibitem [{\citenamefont {Stellinga}\ \emph {et~al.}(2021)\citenamefont {Stellinga}, \citenamefont {Phillips}, \citenamefont {Mekhail}, \citenamefont {Selyem}, \citenamefont {Turtaev}, \citenamefont {{\v{C}}i{\v{z}}m{\'a}r},\ and\ \citenamefont {Padgett}}]{stellinga2021time}%
  \BibitemOpen
  \bibfield  {author} {\bibinfo {author} {\bibfnamefont {D.}~\bibnamefont {Stellinga}}, \bibinfo {author} {\bibfnamefont {D.~B.}\ \bibnamefont {Phillips}}, \bibinfo {author} {\bibfnamefont {S.~P.}\ \bibnamefont {Mekhail}}, \bibinfo {author} {\bibfnamefont {A.}~\bibnamefont {Selyem}}, \bibinfo {author} {\bibfnamefont {S.}~\bibnamefont {Turtaev}}, \bibinfo {author} {\bibfnamefont {T.}~\bibnamefont {{\v{C}}i{\v{z}}m{\'a}r}}, \ and\ \bibinfo {author} {\bibfnamefont {M.~J.}\ \bibnamefont {Padgett}},\ }\bibfield  {title} {\enquote {\bibinfo {title} {Time-of-flight 3d imaging through multimode optical fibers},}\ }\href@noop {} {\bibfield  {journal} {\bibinfo  {journal} {Science}\ }\textbf {\bibinfo {volume} {374}},\ \bibinfo {pages} {1395--1399} (\bibinfo {year} {2021})}\BibitemShut {NoStop}%
\bibitem [{\citenamefont {Dong}\ \emph {et~al.}(2022)\citenamefont {Dong}, \citenamefont {Wen}, \citenamefont {Pang}, \citenamefont {Wang}, \citenamefont {Wu}, \citenamefont {Liu},\ and\ \citenamefont {Yang}}]{dong2022modulated}%
  \BibitemOpen
  \bibfield  {author} {\bibinfo {author} {\bibfnamefont {Z.}~\bibnamefont {Dong}}, \bibinfo {author} {\bibfnamefont {Z.}~\bibnamefont {Wen}}, \bibinfo {author} {\bibfnamefont {C.}~\bibnamefont {Pang}}, \bibinfo {author} {\bibfnamefont {L.}~\bibnamefont {Wang}}, \bibinfo {author} {\bibfnamefont {L.}~\bibnamefont {Wu}}, \bibinfo {author} {\bibfnamefont {X.}~\bibnamefont {Liu}}, \ and\ \bibinfo {author} {\bibfnamefont {Q.}~\bibnamefont {Yang}},\ }\bibfield  {title} {\enquote {\bibinfo {title} {A modulated sparse random matrix for high-resolution and high-speed 3d compressive imaging through a multimode fiber},}\ }\href@noop {} {\bibfield  {journal} {\bibinfo  {journal} {Science bulletin}\ }\textbf {\bibinfo {volume} {67}},\ \bibinfo {pages} {1224--1228} (\bibinfo {year} {2022})}\BibitemShut {NoStop}%
\bibitem [{\citenamefont {Lee}\ \emph {et~al.}(2022)\citenamefont {Lee}, \citenamefont {Parot}, \citenamefont {Bouma},\ and\ \citenamefont {Villiger}}]{lee2022confocal}%
  \BibitemOpen
  \bibfield  {author} {\bibinfo {author} {\bibfnamefont {S.-Y.}\ \bibnamefont {Lee}}, \bibinfo {author} {\bibfnamefont {V.~J.}\ \bibnamefont {Parot}}, \bibinfo {author} {\bibfnamefont {B.~E.}\ \bibnamefont {Bouma}}, \ and\ \bibinfo {author} {\bibfnamefont {M.}~\bibnamefont {Villiger}},\ }\bibfield  {title} {\enquote {\bibinfo {title} {Confocal 3d reflectance imaging through multimode fiber without wavefront shaping},}\ }\href@noop {} {\bibfield  {journal} {\bibinfo  {journal} {Optica}\ }\textbf {\bibinfo {volume} {9}},\ \bibinfo {pages} {112--120} (\bibinfo {year} {2022})}\BibitemShut {NoStop}%
\bibitem [{\citenamefont {Sun}\ \emph {et~al.}(2022)\citenamefont {Sun}, \citenamefont {Wu}, \citenamefont {Wu}, \citenamefont {Goswami}, \citenamefont {Girardo}, \citenamefont {Cao}, \citenamefont {Guck}, \citenamefont {Koukourakis},\ and\ \citenamefont {Czarske}}]{sun2022quantitative}%
  \BibitemOpen
  \bibfield  {author} {\bibinfo {author} {\bibfnamefont {J.}~\bibnamefont {Sun}}, \bibinfo {author} {\bibfnamefont {J.}~\bibnamefont {Wu}}, \bibinfo {author} {\bibfnamefont {S.}~\bibnamefont {Wu}}, \bibinfo {author} {\bibfnamefont {R.}~\bibnamefont {Goswami}}, \bibinfo {author} {\bibfnamefont {S.}~\bibnamefont {Girardo}}, \bibinfo {author} {\bibfnamefont {L.}~\bibnamefont {Cao}}, \bibinfo {author} {\bibfnamefont {J.}~\bibnamefont {Guck}}, \bibinfo {author} {\bibfnamefont {N.}~\bibnamefont {Koukourakis}}, \ and\ \bibinfo {author} {\bibfnamefont {J.~W.}\ \bibnamefont {Czarske}},\ }\bibfield  {title} {\enquote {\bibinfo {title} {Quantitative phase imaging through an ultra-thin lensless fiber endoscope},}\ }\href@noop {} {\bibfield  {journal} {\bibinfo  {journal} {Light: Science \& Applications}\ }\textbf {\bibinfo {volume} {11}},\ \bibinfo {pages} {204} (\bibinfo {year} {2022})}\BibitemShut {NoStop}%
\bibitem [{\citenamefont {Lyu}\ \emph {et~al.}(2023)\citenamefont {Lyu}, \citenamefont {Abrashitova}, \citenamefont {De~Boer}, \citenamefont {Andresen}, \citenamefont {Rigneault},\ and\ \citenamefont {Amitonova}}]{lyu2023sub}%
  \BibitemOpen
  \bibfield  {author} {\bibinfo {author} {\bibfnamefont {Z.}~\bibnamefont {Lyu}}, \bibinfo {author} {\bibfnamefont {K.}~\bibnamefont {Abrashitova}}, \bibinfo {author} {\bibfnamefont {J.~F.}\ \bibnamefont {De~Boer}}, \bibinfo {author} {\bibfnamefont {E.~R.}\ \bibnamefont {Andresen}}, \bibinfo {author} {\bibfnamefont {H.}~\bibnamefont {Rigneault}}, \ and\ \bibinfo {author} {\bibfnamefont {L.~V.}\ \bibnamefont {Amitonova}},\ }\bibfield  {title} {\enquote {\bibinfo {title} {Sub-diffraction computational imaging via a flexible multicore-multimode fiber},}\ }\href@noop {} {\bibfield  {journal} {\bibinfo  {journal} {Optics Express}\ }\textbf {\bibinfo {volume} {31}},\ \bibinfo {pages} {11249--11260} (\bibinfo {year} {2023})}\BibitemShut {NoStop}%
\bibitem [{\citenamefont {Goodman}(2007)}]{goodman2007speckle}%
  \BibitemOpen
  \bibfield  {author} {\bibinfo {author} {\bibfnamefont {J.~W.}\ \bibnamefont {Goodman}},\ }\href@noop {} {\emph {\bibinfo {title} {Speckle phenomena in optics: theory and applications}}}\ (\bibinfo  {publisher} {Roberts and Company Publishers},\ \bibinfo {year} {2007})\BibitemShut {NoStop}%
\bibitem [{\citenamefont {Born}\ and\ \citenamefont {Wolf}(2019)}]{born2019principles}%
  \BibitemOpen
  \bibfield  {author} {\bibinfo {author} {\bibfnamefont {M.}~\bibnamefont {Born}}\ and\ \bibinfo {author} {\bibfnamefont {E.}~\bibnamefont {Wolf}},\ }\href@noop {} {\emph {\bibinfo {title} {Principles of Optics}}}\ (\bibinfo  {publisher} {Cambridge University Press},\ \bibinfo {year} {2019})\BibitemShut {NoStop}%
\bibitem [{\citenamefont {van~den Berg}\ and\ \citenamefont {Friedlander}(2019)}]{spgl1site}%
  \BibitemOpen
  \bibfield  {author} {\bibinfo {author} {\bibfnamefont {E.}~\bibnamefont {van~den Berg}}\ and\ \bibinfo {author} {\bibfnamefont {M.~P.}\ \bibnamefont {Friedlander}},\ }\href@noop {} {\enquote {\bibinfo {title} {{SPGL1}: A solver for large-scale sparse reconstruction},}\ } (\bibinfo {year} {2019}),\ \bibinfo {note} {https://friedlander.io/spgl1}\BibitemShut {NoStop}%
\bibitem [{\citenamefont {van~den Berg}\ and\ \citenamefont {Friedlander}(2008)}]{BergFriedlander:2008}%
  \BibitemOpen
  \bibfield  {author} {\bibinfo {author} {\bibfnamefont {E.}~\bibnamefont {van~den Berg}}\ and\ \bibinfo {author} {\bibfnamefont {M.~P.}\ \bibnamefont {Friedlander}},\ }\bibfield  {title} {\enquote {\bibinfo {title} {Probing the pareto frontier for basis pursuit solutions},}\ }\href {\doibase 10.1137/080714488} {\bibfield  {journal} {\bibinfo  {journal} {SIAM Journal on Scientific Computing}\ }\textbf {\bibinfo {volume} {31}},\ \bibinfo {pages} {890--912} (\bibinfo {year} {2008})}\BibitemShut {NoStop}%
\end{thebibliography}%

\end{document}